\documentstyle[12pt, epsf]{article}
\newcommand\ZZZ{{\hbox{ Z\kern-1.6mm Z}}}
\newcommand{\beq}{\begin{equation}}
\newcommand{\eeq}{\end{equation}}
\newcommand{\bea}{\begin{eqnarray}}
\newcommand{\eea}{\end{eqnarray}}
\newcommand{\ra}{\rangle}
\newcommand{\la}{\langle}
\newcommand{\lt}{\left}
\newcommand{\rt}{\right}
\newcommand{\one}{{\hbox{ 1\kern-1.2mm l}}}

\newcommand{\g}{\gamma}
\newcommand{\gb}{\bar \gamma}
\newcommand{\Z}{{\bf Z}}
\newcommand{\mv}{M^{V}}
\newcommand{\ms}{M^{S}}

\newcommand{\mbs}{\bar M^{S}}

\newcommand{\bps}{\hbox{BPS}}

\newcommand{\al}{\alpha}
\newcommand{\bt}{\beta}
\newcommand{\dt}{\delta}
\newcommand{\alt}{\tilde \alpha}

\newcommand{\N}{{\cal N}}

\newcommand{\del}{\partial}

\newcommand{\zb}{\bar z}
\newcommand{\wb}{\bar w}

\newcommand{\D}{\Delta}

\newcommand{\eps}{\epsilon}
\newcommand{\M}{{\cal M}}

\newcommand{\lam}{\lambda}

\newcommand{\K}{{\cal K}}
\newcommand{\KD}{{\cal K_D}}
\newcommand{\wt}{\tilde w}
\newcommand{\lamt}{\tilde \lambda}
\newcommand{\pt}{\tilde p}
\newcommand{\th}{\theta}
\newcommand{\tht}{\tilde \theta}
\newcommand{\Ut}{\tilde U}
\newcommand{\Ub}{\bar U}
\newcommand{\Vt}{\tilde V}
\newcommand{\Vb}{\bar V}
\newcommand{\tr}{\hbox{Tr}}
\newcommand{\Qt}{\tilde Q}

\newcommand{\sectiono}[1]{\section{#1}\setcounter{equation}{0}}
\newcommand{\subsectiono}[1]{\subsection{#1}}

\begin{document}
{}~
{}~
\hfill\vbox{\hbox{UK/05-03} \hbox{hep-th/0505157}}\break

\vskip .6cm

\centerline{\Large \bf On D-Brane Boundary State Analysis}
\medskip
\centerline{\Large \bf in Pure Spinor Formalism}

\medskip

\vspace*{4.0ex}

\centerline{\large \rm Partha Mukhopadhyay }

\vspace*{4.0ex}

\centerline{\large \it  Department of Physics and Astronomy}

\centerline{\large \it  University of Kentucky, Lexington, KY-40506, U.S.A.}

\medskip

\centerline{E-mail: partha@pa.uky.edu}

\vspace*{5.0ex}

\centerline{\bf Abstract} \bigskip

We explore a particular approach to study D-brane boundary states
in Berkovits' pure spinor formalism of superstring theories. In
this approach one constructs the boundary states in the relevant
conformal field theory by relaxing the pure spinor constraints.
This enables us to write down the open string boundary conditions for
non-BPS D-branes in type II string theories, generalizing our previous
work in light-cone Green-Schwarz formalism. As a first step to explore
how to apply these boundary states for physical computations we
prescribe rules for computing disk one point functions for the
supergravity modes. We also comment on the force computation between
two D-branes and point out that it is hard to make the world-sheet
open-closed duality manifest in this computation.

\vfill \eject

\tableofcontents

\baselineskip=18pt

\sectiono{Introduction and Summary} \label{s:intro}

Quantizing superstrings with manifest super-Poincar$\acute{\hbox{e}}$
covariance is a
long standing problem. Several years ago an extension of Siegel's approach
\cite{siegel85} to Green-Schwarz (GS) superstrings \cite{green83} was
proposed by Berkovits \cite{purespinor, cohomology, scattering-tree,
scattering-loop, PS-NSR-GS}, where a set of bosonic ghost degrees of
freedom, which are space-time spinors satisfying a pure spinor
constraint, were introduced. Various intuitive rules have been
suggested for necessary computations. It has been argued
\cite{cohomology} that it gives the correct physical perturbative
spectrum of string theory. Rules for computing scattering amplitudes
have also been formulated \cite{purespinor, scattering-tree,
scattering-loop} which have produced super-Poincar$\acute{\hbox{e}}$
covariant results more easily than the standard Nevew-Schwarz-Ramond
(NSR) formalism. These
striking results make it essential to study this approach deeply in
order to get new insights into superstring theories\footnote{See
 \cite{other} for other relevant works and \cite{backgrounds} for
studies on different non-trivial backgrounds.}. Certainly an important
issue to be considered is the study of D-branes in this context. This
has been done from various points of view in \cite{purespinorBI,
anguelova03, schiappa05} (see also \cite{howe05}). In particular,
in \cite{schiappa05} Schiappa
and Wyllard studied the boundary state analysis for D-branes and disk
scattering amplitudes. Due to the pure spinor constraint the
construction of boundary states, which involved relating the variables
of pure spinor and NSR formalisms \cite{PS-NSR-GS}, looks
complicated. To avoid such complications we shall explore an
alternative approach which might provide a convenient computational
tool for analysis involving boundary states. We shall discuss this
below after going through some basic relevant features of pure spinor
formalism.

In pure spinor formalism one starts out with a given conformal field
theory (CFT) which is supposed to be the conformal gauge fixed form of
a local action\footnote{Attempts have been made in \cite{origin} to understand
the origin of this approach.}. In addition to the usual bosonic matter
of the standard NSR formalism, this CFT also includes a fermionic
matter and a bosonic ghost parts both of which contain space-time
fermions. All the world-sheet fields are free except that the ghost
fields are required to satisfy a covariant pure spinor
constraint\footnote{Extensions of the present framework by relaxing
this constraint have been considered in \cite{grassi, aisaka}.} which
removes the desired number of degrees of freedom. Naturally, this
constraint makes the Hilbert space structure of the theory more
complicated than the one corresponding to the unconstrained CFT. The
physical states are given by the states of certain ghost number in the
cohomology of a proposed BRST operator \cite{purespinor}. As usual,
there exists linearized gauge transformation provided by the BRST
exact states. Although it is understood how the massless closed string
states arise in this formalism, arbitrarily high massive states are
difficult to describe. This is because of the pure spinor constraint
and the fact that, to begin with, the vertex operators are arbitrary
functions on the $d=9+1$, ${\cal N}=1$ superspace (for open strings).
Although the latter may be expected for a manifestly
super-Poincar$\acute{\hbox{e}}$
covariant formalism, it results in a lot of redundant
fields \cite{sugra, massive, grassi04} which need to be removed by gauge
fixation, a procedure that has to be done separately at every level. In NSR
formalism the associated gauge symmetry is fixed by a set of well known simple
conditions. Moreover, in this gauge the quadratic part of the
space-time action simplifies in a certain manner so that the
propagator
gets an interpretation of the world-sheet time evolution \cite{witten}.
This sits at the heart of the fact that world-sheet open-closed
duality is manifest in NSR computations. It is not yet clear what
should be the analog of this gauge choice in pure spinor formalism.

Let us now discuss D-brane boundary states in this
context\footnote{See, for example, \cite{NSRbs} for reviews on this
subject in NSR formalism.}.
Given a closed string Hilbert space this state is found as a
solution to the open string boundary condition expressed in the closed
string channel. Construction of the Hilbert space covariantly in pure
spinor formalism is not very easy as the CFT is actually interacting
because of the pure spinor constraint. Nevertheless, since the
interaction is introduced only through the constraint, one might
expect that this Hilbert space can be embedded in the bigger Hilbert
space of the unconstrained theory which is completely free. Our
approach will be to construct boundary states in the unconstrained
theory and to prescribe rules for computing physical quantities using
them. Computationally, this may prove favorable provided all such
rules can be consistently set up. This approach makes it easy to write
down the open string boundary conditions and boundary states for BPS
D-branes. The same for non-BPS D-branes in light-cone GS formalism
were recently found in \cite{mukhopadhyay04}. The open string boundary
conditions turned out to relate bi-local operators that are quadratic
in space-time fermions. We shall see that it has a simple
generalization to the present case.

To explore exactly how these boundary states should be used for
physical computations we first consider computing the strength of the
closed string sources that D-branes produce.  Here we demonstrate that
indeed a consistent set of rules can be prescribed at least for the
supergravity modes, leaving its generalization to higher massive modes
and to arbitrary disk scattering amplitudes for future work. Another
interesting computation to be understood is the so-called {\it cylinder
diagram} which gives the force between two D-branes and demonstrates
the world-sheet open-closed duality. Because of the problem of gauge
fixation mentioned earlier, this computation is not straightforward in
pure spinor formalism. Moreover, in NSR formalism ghosts produce a
background independent contribution which cancels two (light-cone)
coordinates worth of contribution from the matter part. We emphasize
that it is difficult to get such background independent contribution
in pure spinor formalism. We demonstrate this with the simplest
computation, namely the long range NS-NS force between two parallel
D-branes where only the massless NS-NS states are involved. The
relevant ghost contribution in NSR formalism comes from the
ghost-dilaton which does not have any analog in pure spinor formalism.
This does not necessarily imply any inconsistency as the whole computation
can be performed in supergravity where one is only required to
evaluate a tree-level Feynman diagram between two
sources \cite{NSRreview}. The only role played by the boundary states
in this computation is to provide the correct strength for the
sources\footnote{This statement is true even at the full string theoretic
level as long as there is a space-time field theoretic way of
computing the force.}. Obviously the world-sheet open-closed duality
will not be manifest in such a computation. Extending this argument to
higher massive levels we suggest that the present boundary state
should actually be compared with the NSR boundary states in old
covariant quantization. Computation of the R-R amplitude is complicated even in
the NSR formalism because of the superghost zero modes
\cite{sgh0}. Its study in the pure spinor formalism may require
special attention which we leave for future work.

The rest of the paper is organized as follows. We review the basic CFT
structure of the pure spinor formalism in sec.\ref{s:CFT}. The BPS and
non-BPS boundary conditions and boundary states have been discussed in
sec.\ref{s:bcbs}. Sec.\ref{s:physical} concentrates on physical
computations where we demonstrate how to compute disk one point
functions for the supergravity modes and comment on the force
computation. Sec.\ref{s:discussion} discusses some future directions.
Our convention for the gamma matrices and relevant identities are
given in the appendix.

\sectiono{The Conformal Field Theory}
\label{s:CFT}

In pure spinor formalism one begins with a CFT which has the following
three parts for type II string theories \cite{purespinor}:
\bea
S = S_B + S_F + S_G ~,
\label{S}
\eea
where $S_B$ is same as the bosonic matter conformal field theory in
the usual NSR formalism and therefore has central charge $c_B =\tilde
c_B =10$. The left moving part of $S_F$ is the direct sum of $16$
fermionic $(b,c)$ theories, namely $(p_{\al}(z),~\th^{\al}(z))$ where
$\al =1,2, \cdots ,16$ is a space-time spinor index that has been
explained in appendix \ref{a:gamma}. Conformal dimensions of the
fields are as follows: $h_{p_{\al}}=1,~ h_{\th_{\al}}=0$. For type IIB
string theory the right moving part is
same as the left moving part whereas for type IIA the space-time
chirality of the fields are just opposite,
i.e. $(p^{\al}(\zb),~\th_{\al}(\zb))$. Therefore the central charges
for $S_F$ are $c_F=\tilde c_F=-32$. The bosonic ghost part $S_G$ is
the most difficult part. The left moving part of it is given by the
direct sum of $16$ bosonic $(\bt, \g)$ systems:
$(w_{\al}(z),\lam^{\al}(z))$ with conformal dimensions
$h_{w_{\al}}=1,~h_{\lam_{\al}}=0$ and with a pure spinor constraint on
$\lam^{\al}$. Just like the case of $S_F$ the right moving part of
$S_G$ is same as the left moving part for type IIB and the fields take
opposite space-time chirality for type IIA. For definiteness we shall
hereafter consider only type IIB string theory. The pure spinor
constraints are given by,
\bea
\lam^{\al}(z)\gb^{\mu}_{\al \bt} \lam^{\bt}(z) =
\lamt^{\al}(\zb)\gb^{\mu}_{\al \bt} \lamt^{\bt}(\zb) = 0~.
\label{PSconstraint}
\eea
Our notation for the gamma matrices can be found in appendix
\ref{a:gamma}.
Because of these pure spinor constraints there are actually $11$
independent fields instead of $16$ and therefore $S_G$ has central
charges $c_G=\tilde c_G=22$ (see \cite{character} for a recent
covariant computation of this central charge), instead of $32$. This
makes the total
central charge of $S$ zero. The local gauge transformations
corresponding to the above constraints are given by,
\bea
\dt \lam^{\al}(z) = 0~, & \quad & \dt \tilde \lam^{\al}(\zb) =0 ~, \cr
\dt w_{\al}(z) = \Lambda_{\mu}(z) \gb^{\mu}_{\al \bt} \lam^{\bt}(z)~,
&\quad &
\dt \wt_{\al}(\zb) = \tilde \Lambda_{\mu}(\zb) \gb^{\mu}_{\al \bt}
\lamt^{\bt}(\zb)~,
\label{dtw}
\eea
which reduce the degrees of freedom of $w_{\al}$ and $\tilde
w_{\al}$. As a result only the gauge invariant operators
$\lam^{\al}(z)w_{\al}(z)$ and $\lam^{\al}(z)\gb^{\mu \nu~\bt}_{~~\al}
w_{\bt}(z)$ (and similarly for the right moving sector) can appear in
the construction of physical states.
Space-time supersymmetry and BRST currents are given by ($\alpha^{\prime}=2$),
\bea
q_{\al} &=& p_{\al} +{1\over 2} (\gb^{\mu} \th)_{\al}
\del X_{\mu} + {1\over 24} (\gb^{\mu}\th)_{\al} (\th \gb \del \th)~, \cr
j_B &=& \lam^{\al} d_{\al}~, \quad d_{\al}= p_{\al} - {1\over 2}
(\gb^{\mu}\th)_{\al} \del X_{\mu} - {1\over 8} (\gb^{\mu}\th)_{\al}
(\th \gb_{\mu} \del \th) ~,
\label{qal-jB}
\eea
and similarly for the right moving components.

Covariant quantization of the above CFT is not straightforward due to
the pure spinor constraint. Nevertheless all the allowed states should
form a subspace of the unconstrained Hilbert space which corresponds
to the completely free theory. Performing the usual mode expansion
with periodic boundary conditions for the world-sheet fields and
quantizing the free theory one gets the following nontrivial commutation relations,
\bea
[\lam^{\al}_m, w_{\bt ,n}] =
\dt^{\al}_{~\bt} \dt_{m+n}~, \quad \{\th^{\al}_m, p_{\bt, n}\} =
\dt^{\al}_{~\bt} \dt_{m+n}~,  \quad m,n \in \Z~,
\label{lam-th-commut}
\eea
and similarly for the right moving sector. The full Hilbert space is
obtained by applying the negative modes freely on the ground
states. Defining the states $|0\ra$ and $|\hat 0\ra$ in the following
way,
\bea
\lt. \begin{array}{l}
\lam^{\al}_n \\ \\ \th^{\al}_n
\end{array} \rt \} |0\ra =0 ~, ~\forall \al, n> 0~, &\quad &
\lt. \begin{array}{l}
w_{\al ,n} \\ \\ p_{\al ,n}
\end{array} \rt \} |0\ra =0 ~, ~\forall \al, n\geq 0~, \cr && \cr && \cr
\lt. \begin{array}{l}
\lam^{\al}_n \\ \\ \th^{\al}_n
\end{array} \rt \} |\hat 0\ra =0 ~, ~\forall \al, n \geq 0~, &\quad &
\lt. \begin{array}{l}
w_{\al ,n} \\ \\ p_{\al ,n}
\end{array} \rt \} |\hat 0\ra =0 ~, ~\forall \al, n> 0~,
\label{zero-zerohat}
\eea
all the ground states having the same $L_0$ eigenvalue can be obtained
by either applying
$\lam^{\al}_0$ and $\th^{\al}_0$ repeatedly on $|0\ra$ or applying
$w_{\al,0}$ and $p_{\al,0}$ repeatedly on $|\hat 0\ra$. Since the
ghost sector is bosonic
the number of ground states corresponding to this sector is actually
infinite. The ground states of the $(p_{\al},\th^{\al})$ system, along
with those of the bosonic matter, construct the $d=9+1,~{\cal N}=1$
superspace.

Assigning the ghost numbers $(g, \tilde g)$ to various fields in the
following way: $\lam^{\al} \to (1, 0)$, $w_{\al} \to (-1,0)$,
$\lamt^{\al} \to (0,1)$, $\wt_{\al} \to (0,-1)$ and $\hbox{others} \to
(0,0)$, a  physical on-shell vertex operator $V(z, \zb)$ is defined to
be a $(1,1)$ operator such that,
\bea
Q_B | V \ra = 0~, \quad \Qt_B |V \ra =0~,
\label{onshell}
\eea
where $Q_B= \displaystyle{\oint {dz \over 2\pi i} j_B(z)}$ (similarly
for $\tilde Q_B$) and $|V\ra = V(0,0) |0\ra \otimes
\widetilde{|0\ra}$. Clearly the linearized gauge transformation is
given by, $|\dt V\ra = Q_B |\Phi \ra + \Qt_B |\tilde \Phi \ra$ for any
ghost number $(0,1)$ and $(1,0)$ operators $\Phi $ and $\tilde \Phi $
respectively. In particular, the massless vertex operators are given
by,
\bea
\hbox{NS-NS :} && a^{(\mu}(z)\tilde a^{\nu)}(\zb) e^{ik.X}(z,\zb)~, \cr
\hbox{NS-R :} && a^{\mu}(z) \tilde \chi_{\al}(\zb) e^{ik.X}(z,\zb)~, \cr
\hbox{R-NS :} && \chi_{\al}(z) \tilde a^{\mu}(\zb) e^{ik.X}(z,\zb)~, \cr
\hbox{R-R (field strength) :}&& \chi_{\al}(z) \tilde \chi_{\bt}(\zb)
e^{ik.X}(z,\zb)~,
\label{Vop}
\eea
where, to the lowest order in the $\theta$-expansion
\cite{scattering-tree}\footnote{Although the vertex operators are BRST
invariant only when the full $\theta$-expansions are considered, the leading
order terms will suffice for our computations. },
\bea
a^{\mu}(z)=\lam(z) \gb^{\mu} \th(z)~, \quad
\chi_{\al}(z)= (\lam(z)\gb^{\mu}\th(z))(\gb_{\mu}\th(z))_{\al}~,
\eea
and similarly for the right moving operators.
The zero modes saturation rule suggested by Berkovits
\cite{purespinor} can be obtained by choosing a particular out-going
ground state in the unconstrained theory \cite{chesterman04},
\bea
&& \la (\lam_0 \gb^{\mu} \th_0)(\lam_0 \gb^{\nu} \th_0)(\lam_0
\gb^{\rho} \th_0) (\th_0 \gb_{\mu \nu \rho} \th_0) \ra_{Berkovits} \cr
&&= \la \Omega | (\lam_0 \gb^{\mu} \th_0)(\lam_0 \gb^{\nu}
\th_0)(\lam_0 \gb^{\rho} \th_0) (\th_0 \gb_{\mu \nu \rho} \th_0) |0\ra
= 1~,
\label{c}
\eea
where,
\bea
|\Omega \ra = {1\over c}
(w_0\g^{\mu}p_0)(w_0\g^{\nu}p_0)(w_0\g^{\rho}p_0)(p_0\g_{\mu \nu
\rho}p_0)|\hat 0 \ra~,
\label{Omega}
\eea
$c$ being a numerical constant chosen properly to satisfy the second
line of eq.(\ref{c}). Here we adopt the following notation: $\la \cdots \ra_{Berkovits}$ refers to a correlation function computed in the actual constrained CFT. Any other inner product will be computed in the free CFT.
Notice that according to the convention of Chesterman
\cite{chesterman04}, the ghost numbers for the states $|0\ra$ and
$|\hat 0\ra$ are $8$ and $-8$ respectively. Following the same
convention we can find the ghost number of any given state in this
CFT.

\sectiono{Boundary Conditions and Boundary States}
\label{s:bcbs}

Here we concentrate only on the combined fermionic matter and bosonic
ghost part of the CFT as the bosonic matter part is well
understood. As has been discussed before, we shall study the open
string boundary conditions and the corresponding boundary states in
the unconstrained
CFT\footnote{Although the unconstrained CFT is not relevant to any
string theory and therefore there does not exist any open string
interpretation, we shall still continue to use this terminology by the
abuse of language.}. The traditional method of finding these boundary
conditions is to take variation of the world-sheet action with respect
to the basic fields and then set the boundary term to zero. In our
case these conditions give the following equations on the upper half
plane (UHP),
\bea
\lt.
\begin{array}{c}
w_{\al}(z) \dt \lam^{\al}(z) = \wt_{\al}(\zb) \dt \lamt^{\al}(\zb)~,
\\ \\
p_{\al}(z) \dt \th^{\al}(z) = \pt_{\al}(\zb) \dt \tht^{\al}(\zb)~,
\end{array} \rt\} \hbox{at } z=\zb~.
\label{b-term}
\eea
As was pointed out in \cite{mukhopadhyay04}, although the BPS boundary conditions
can be easily obtained from the above conditions, the ones corresponding to
non-BPS D-branes are not straightforward. We shall discuss both the
cases below.

\subsectiono{BPS D-Branes}
\label{s:bcbsBPS}

Let us consider a type IIB BPS D$p$-brane ($p =$ odd) aligned
along $x^0, x^1, \cdots, x^p$. Introducing the column vectors, \bea
U^{\al}(z) = \pmatrix{\lam^{\al}(z) \cr \th^{\al}(z)}~, ~~
V_{\al}(z) = \pmatrix{w_{\al}(z) \cr p_{\al}(z)}~, \label{UV} \eea
and similarly for the right moving sector, the open string boundary
conditions which satisfy eqs. (\ref{b-term}) can be written as,
\bea
U^{\al}(z) = \eta (\ms)^{\al}_{~\bt} \Ut^{\bt}(\zb)~, ~~ V_{\al}(z)
= -\eta (\mbs)_{\al}^{~\bt} \Vt_{\bt}(\zb) ~, ~~\hbox{at } z=\zb~,
\label{bcBPS}
\eea
where $\eta =\pm 1$ which correspond to brane and anti-brane
respectively. The spinor matrices represent a set of reflections
along the Neumann directions in the following way,
\bea
\ms = \g^{01\cdots p}~, \quad  \mbs = \gb^{01\cdots p}~.
\label{msmbs}
\eea
The multi-indexed gamma matrices are defined in appendix \ref{a:gamma}.
For the lorentzian D-branes that we are considering here,
\bea
\mbs(\ms)^T = (\mbs)^T\ms = -\one_{16}~.
\label{mbsmsT}
\eea
These matrices also satisfy the following relations,
\bea
\ms \g^{\mu} (\ms)^T = -(\mv)^{\mu}_{~\nu} \g^{\nu}~, &&
(\ms)^T \gb^{\mu} \ms = -(\mv)^{\mu}_{~\nu} \gb^{\nu}~, \cr
(\mbs)^T \g^{\mu} \mbs = -(\mv)^{\mu}_{~\nu} \g^{\nu}~, &&
\mbs \gb^{\mu} (\mbs)^T = -(\mv)^{\mu}_{~\nu} \gb^{\nu}~,
\label{relmsmbs}
\eea
where
$\mv$ is the vector representation of the set of reflections along the
Neumann directions,
\bea
(\mv)^{\mu}_{~\nu} =
\eps_{(\mu)} \dt^{\mu}_{~\nu}~, \quad \eps_{(\mu)}=\lt\{
\begin{array}{ll} -1 & \mu = 0,1,\cdots ,p, \\
+1 & \mu = p+1, \cdots ,9~,
\end{array} \rt.
\label{mv}
\eea
The boundary conditions for the supersymmetry and BRST currents turn out to be,
\bea
q_{\al}(z) = -\eta (\mbs)_{\al}^{~\bt} \tilde q_{\bt}(\zb) ~, \quad
j_B(z) = \tilde j_B(\zb) ~, \quad \hbox{at } z=\zb~.
\label{qjBbc}
\eea
The boundary state for such a D-brane situated at the origin of
the transverse directions is given by,
\bea
|\bps,p,\eta \ra = \N_p
\int \vec dk_{\perp} \exp \lt( \sum_{n\geq 1} {1\over n} \al_{\mu,-n}
(\mv)^{\mu}_{~\nu} \alt^{\nu}_{-n} \rt) |\vec k_{\perp}\ra \otimes
|F,p,\eta \ra~, \label{bsBPS}
\eea
where $\N_p$ is a normalization constant proportional to the D-brane tension, $|\vec k_{\perp}\ra$ is the bosonic Foch vacuum with momentum $\vec k_{\perp}$ along the transverse directions, the exponential factor is the usual bosonic oscillator part \cite{NSRbs} and  $|F,p,\eta \ra $ is the combined fermionic matter and
ghost part which satisfies the following closed string gluing
conditions obtained from the boundary conditions (\ref{bcBPS}),
\bea \lt.
\begin{array}{l} U^{\al}_n -\eta (\ms)^{\al}_{~\bt} \Ut^{\bt}_{-n}
\\  \\ V_{\al,n}-\eta (\mbs)_{\al}^{~\bt} \Vt_{\bt,-n} \end{array} \rt\}
|F,p,\eta \ra = 0 ~, \quad \forall n \in \Z~.
\label{gluingBPS}
\eea
The solution is given by,
\bea
|F,p,\eta \ra &=& \exp \lt[ -\eta \sum_{n\geq 1} \lt( U^{\al T}_{-n}
(\mbs)_{\al}^{~\bt} \sigma_3 \Vt_{\bt, -n} -
\Ut^{\al T}_{-n} (M^{S T})_{\al}^{~\bt} \sigma_3 V_{\bt,-n} \rt)\rt]
|F,p,\eta \ra_0 ~, \cr &&
\label{Fpeta}
\eea
where $\sigma_3= \hbox{diag}(1,-1)$ is the third Pauli matrix and the
zero modes part
$|F,p,\eta \ra_0$ can be given the following two different forms:
\bea
|F,p,\eta \ra_0 &=& \exp \lt( -\eta U^{\al T}_0 (\mbs)_{\al}^{~\bt} \sigma_3
\Vt_{\bt,0} \rt) |0\ra \otimes \widetilde{|\hat 0 \ra}~, \quad \quad
\hbox{I}~, \cr && \cr
&=& \exp \lt( \eta \Ut^{\al T}_0 (M^{S^T})_{\al}^{~\bt} \sigma_3
V_{\bt,0} \rt) |\hat 0\ra \otimes \widetilde{|0 \ra}~, \quad \quad \hbox{II}~.
\label{Fpeta0}
\eea
We shall see in section \ref{ss:one-point} that both the above two
forms can be used to compute one-point functions of the supergravity
modes.

\subsectiono{Non-BPS D-Branes}
\label{ss:bcbsnonBPS}

Non-BPS D-branes in light-cone Green-Schwarz formalism have been
studied in \cite{mukhopadhyay00, nemani, mukhopadhyay04}. In
particular, the covariant open string boundary conditions were found
in \cite{mukhopadhyay04}. Generalizing this work
to any manifestly supersymmetric formalism we may write down the following
general rules for finding the non-BPS boundary conditions on UHP,
\begin{enumerate}
\item
Given a set of all the left moving world-sheet fields that are in
space-time spinor
representation, pair them up in all possible ways to form bi-local
operators such as $A^{\al}(z) B^{\bt}(w)$, $A^{\al}(z)B_{\bt}(w)$,
$A_{\al}(z)B^{\bt}(w)$ or $A_{\al}(z)B_{\bt}(w)$.
\item
Expand them using Fiertz identities summarized in eqs.(\ref{fiertz}).
These expansions will contain bi-local operators in tensor
representations only\footnote{As
pointed out in \cite{mukhopadhyay04}, there is a subtlety involving the
self-dual tensor operator in this expansion which needs to be taken
care of. We shall do this explicitly below.}.
\item
Equate the left and right moving bi-local operators in tensor
representations on the real line by twisting them by the
reflection  matrix in vector representation.
\end{enumerate}
In the present case we use the above rules to obtain the following
open string boundary conditions in the unconstrained CFT:
\bea
\lt. \begin{array}{l}
U^{\al}(z)U^{\bt T}(w) = \M^{\al \bt}_{\g \dt}
\Ut^{\g}(\zb) \Ut^{\dt T}(\wb)~, \\ \\
U^{\al}(z)V_{\bt}^T(w) = \M^{\al~~\dt}_{~\bt \g}
\Ut^{\g}(\zb) \Vt_{\dt}^T(\wb)~, \\ \\
V_{\al}(z)V_{\bt}^T(w) = \M^{\g \dt}_{\al \bt}
\Vt_{\g}(\zb) \Vt_{\dt}^T(\wb)~,
\end{array} \rt\} \hbox{at } z=\zb, w=\wb ~,
\label{bcnBPS} \eea
where the coupling matrices are given by (the relevant Fiertz
identities are listed in appendix \ref{a:gamma}),
\bea
\M^{\al \bt}_{\g \dt} &=& -\lt[ {1\over 16} \g_{\mu}^{\al \bt}
(\mv)^{\mu}_{~\nu} \gb^{\nu}_{\g \dt} + {1\over 16\times 3!}
\g_{\mu_1 \cdots \mu_3}^{\al \bt} (\mv)^{\mu_1}_{~\nu_1}\cdots
(\mv)^{\mu_3}_{~\nu_3} \gb^{\nu_1\cdots \nu_3}_{\g \dt} \rt. \cr &&
\lt. +
{1\over 16\times 5!} \sum_{\mu_1,\cdots ,\mu_5 \in \K}
\g_{\mu_1\cdots \mu_5}^{\al \bt} (\mv)^{\mu_1}_{~\nu_1} \cdots
(\mv)^{\mu_5}_{~\nu_5} \gb^{\nu_1\cdots \nu_5}_{\g \dt} \rt]~, \cr
\M^{\al~~\dt}_{~\bt \g} &=& {1\over 16} \dt^{\al}_{~\bt}
\dt_{\g}^{~\dt} + {1\over 16 \times 2!}
\g_{\mu_1\mu_2~\bt}^{~~~~\al} (\mv)^{\mu_1}_{~\nu_1}
(\mv)^{\mu_2}_{~\nu_2} \gb^{\nu_1\nu_2~\dt}_{~~~~\g} \cr && +{1\over
16 \times 4!} \g_{\mu_1\cdots \mu_4~\bt}^{~~~~~~\al}
(\mv)^{\mu_1}_{~\nu_1} \cdots (\mv)^{\mu_4}_{~\nu_4}
\gb^{\nu_1\cdots \nu_4~\dt}_{~~~~~~\g}~.
\label{calMs}
\eea
The summation convention for the repeated indices has been followed
for all the terms in the above two equations except for the last term
of the first equation. The sum over the five vector indices $\mu_1
\cdots \mu_5$ has been explicitly restricted to a set $\K$ which is
defined as follows. We divide the set of all possible sets of five
indices $\{\{\mu_1,\cdots,\mu_5\}|\mu_i =0,\cdots ,9\}$ into two
subsets of equal order, namely $\K$ and $\KD$ such that for every
element $\{\mu_1,\cdots,\mu_5\} \in \K$ there exists a dual element
$\{\mu_1,\cdots,\mu_5\}_D=\{\nu_1,\cdots,\nu_5\} \in \KD$ such that,
$\eps^{\mu_1\cdots \mu_5\nu_1\cdots \nu_5} \neq 0$. A similar restriction
is intimately related to the basis construction in light-cone
Green-Schwarz formalism \cite{mukhopadhyay04}. Using the properties
(\ref{Poincare-dual})
one can argue that replacing the restricted summation by a free
summation would lead to zero for that particular term when $\mv$
corresponds to a non-BPS D-brane. In that case eqs.(\ref{bcnBPS})
will not be invertible. As argued in \cite{mukhopadhyay04}, since the boundary
conditions (\ref{bcnBPS}) are bi-local, one can take variation of fields
independently at the two points. Using this one can easily show that
eqs.(\ref{b-term}) are satisfied. The supersymmetry current in
eq.(\ref{qal-jB}), being space-time fermionic, do not satisfy a
linear boundary condition. By using covariance one can argue that it
satisfies the same boundary condition as $p_{\alpha}$ which can be
read out from the last equation in (\ref{bcnBPS}). Absence of a
linear boundary condition such as in eq.(\ref{qjBbc}) implies that all
the supersymmetries are broken. But the BRST current does satisfy the same
condition as in eq.(\ref{qjBbc}).

It may seem difficult to obtain the boundary states corresponding to the
above boundary conditions. But we follow a trick discussed in
\cite{mukhopadhyay04} to achieve this. By a "bosonization and
refermionization" method we first define a new set of right moving
fields $\Ub^{\al}(\zb)$ and $\Vb_{\al}(\zb)$ in the following way,
\bea
\Ub^{\al}(\zb) \gb^{\mu}_{\al \bt} \Ub^{\bt T}(\wb) &=&
- (\mv)^{\mu}_{~\nu} \Ut^{\al}(\zb) \gb^{\nu}_{\al \bt} \Ut^{\bt
T}(\wb) ~, \cr && \cr
\Ub^{\al}(\zb) \gb^{\mu_1\cdots \mu_3}_{\al
\bt} \Ub^{\bt T}(\wb) &=& - (\mv)^{\mu_1}_{~\nu_1} \cdots
(\mv)^{\mu_3}_{~\nu_3} \Ut^{\al}(\zb) \gb^{\nu_1\cdots \nu_3}_{\al
\bt} \Ut^{\bt T}(\wb) ~, \cr && \cr
\Ub^{\al}(\zb) \gb^{\mu_1\cdots
\mu_5}_{\al \bt} \Ub^{\bt T}(\wb) &=& - \lt\{ \begin{array}{rl}
(\mv)^{\mu_1}_{~\nu_1} \cdots (\mv)^{\mu_5}_{~\nu_5} \Ut^{\al}(\zb)
\gb^{\nu_1\cdots \nu_5}_{\al \bt} \Ut^{\bt T}(\wb)~,
& \mu_1, \cdots , \mu_5 \in \K ~, \\ \\
-(\mv)^{\mu_1}_{~\nu_1} \cdots (\mv)^{\mu_5}_{~\nu_5}
\Ut^{\al}(\zb) \gb^{\nu_1\cdots \nu_5}_{\al \bt} \Ut^{\bt T}(\wb)~,
& \mu_1, \cdots , \mu_5 \in \KD ~, \end{array} \rt. \cr &&
\label{UbUb}
\eea

\bea
\Vb_{\al}(\zb) \g^{\mu \al \bt} \Vb_{\bt}^T(\wb) &=&
- (\mv)^{\mu}_{~\nu} \Vt_{\al}(\zb) \g^{\nu \al \bt}
\Vt_{\bt}^T(\wb) ~, \cr && \cr
\Vb_{\al}(\zb) \g^{\mu_1\cdots \mu_3 \al \bt}
\Vb_{\bt}^T(\wb) &=&
- (\mv)^{\mu_1}_{~\nu_1} \cdots (\mv)^{\mu_3}_{~\nu_3}
\Vt_{\al}(\zb) \gb^{\nu_1\cdots \nu_3 \al \bt}
\Vt_{\bt}^T(\wb) ~, \cr && \cr
\Vb_{\al}(\zb) \g^{\mu_1\cdots \mu_5 \al \bt}
\Vb_{\bt}^T(\wb) &=& - \lt\{ \begin{array}{rl}
(\mv)^{\mu_1}_{~\nu_1} \cdots (\mv)^{\mu_5}_{~\nu_5}
\Vt_{\al}(\zb) \g^{\nu_1\cdots \nu_5 \al \bt} \Vt_{\bt}^T(\wb)~,
& \mu_1, \cdots , \mu_5 \in \K ~, \\ \\
-(\mv)^{\mu_1}_{~\nu_1} \cdots (\mv)^{\mu_5}_{~\nu_5}
\Vt_{\al}(\zb) \g^{\nu_1\cdots \nu_5 \al \bt} \Vt_{\bt}^T(\wb)~,
& \mu_1, \cdots , \mu_5 \in \KD ~, \end{array} \rt. \cr &&
\label{VbVb}
\eea

\bea
\Ub^{\al}(\zb) \Vb^T_{\al}(\wb) &=& \Ut^{\al}(\zb) \Vt^T_{\al}(\wb)
~, \cr
\Ub^{\al}(\zb) \gb^{\mu_1\mu_2~\bt}_{~~~~\al} \Vb^T_{\bt}(\wb) &=&
(\mv)^{\mu_1}_{~\nu_1} (\mv)^{\mu_2}_{~\nu_2}
\Ut^{\al}(\zb) \gb^{\nu_1\nu_2~\bt}_{~~~~\al} \Vt^T_{\bt}(\wb) ~, \cr
\Ub^{\al}(\zb) \gb^{\mu_1\cdots \mu_4~\bt}_{~~~~~~\al} \Vb^T_{\bt}(\wb) &=&
(\mv)^{\mu_1}_{~\nu_1}\cdots  (\mv)^{\mu_4}_{~\nu_4}
\Ut^{\al}(\zb) \gb^{\nu_1\cdots \nu_4~\bt}_{~~~~~~\al} \Vt^T_{\bt}(\wb) ~.
\label{UbVb}
\eea
Notice the definitions of the anti-self-dual and self-dual operators
$\Ub^{\al}(\zb) \gb^{\mu_1\cdots \mu_5}_{\al \bt} \Ub^{\bt T}(\wb)$
and $\Vb_{\al}(\zb) \g^{\mu_1\cdots \mu_5 \al \bt} \Vb_{\bt}^T(\wb)$
in eqs.(\ref{UbUb}) and (\ref{VbVb}) respectively. There is a sign
difference between the cases when the set of indices belong to $\K$
and $\KD$. This is because for a non-BPS D-brane $\mv$ represents a
set of odd number of reflections under which a self-dual tensor
transforms to an anti-self-dual tensor and vice-versa.
In terms of these new fields the boundary conditions in (\ref{bcnBPS})
take the following simpler form,
\bea
\lt. \begin{array}{l}
U^{\al}(z)U^{\bt T}(w) =\Ub^{\al}(\zb) \Ub^{\bt T}(\wb)~, \\ \\
U^{\al}(z)V_{\bt}^T(w) =\Ub^{\al}(\zb) \Vb_{\bt}^T(\wb)~, \\ \\
V_{\al}(z)V_{\bt}^T(w) =\Vb_{\al}(\zb) \Vb_{\bt}^T(\wb)~,
\end{array} \rt\} \hbox{at } z=\zb, w=\wb ~.
\label{bcnBPSbar}
\eea
The closed string gluing conditions obtained from these boundary
conditions are given by,
\bea
\lt. \begin{array}{r}
(U^{\al}_m U^{\bt T}_n - \Ub^{\al}_{-m} \Ub^{\bt T}_{-n})\\ \\
(U^{\al}_m V_{\bt,n}^T + \Ub^{\al}_{-m} \Vb_{\bt,-n}^T)\\ \\
(V_{\al, m} V_{\bt,n}^T - \Vb_{\al, -m} \Vb_{\bt, -n}^T)
\end{array} \rt\} |F,p\ra =0 ~, \quad \forall m,n \in \Z~.
\eea
One can show that this is precisely the gluing conditions satisfied by
the NS-NS part of the D$9$ boundary state, which is given by
eqs.(\ref{Fpeta}, \ref{Fpeta0}) for $\ms = \one_{16}$ and $\mbs
=-\one_{16}$, with the right moving oscillators replaced by the
corresponding barred oscillators. Notice
that there is no replacement for the right moving ground state in either
$|0\ra \otimes \widetilde{|\hat 0\ra}$ or $|\hat 0\ra\otimes \widetilde{|0\ra}$. This is because
$\widetilde{|0\ra} $ and $\widetilde{|\hat 0\ra}$ are the two ground
states that satisfy the same equations as (\ref{zero-zerohat}) with
all the oscillators replaced by the corresponding barred
oscillators. Since it is the NS-NS part of the state in (\ref{Fpeta},
\ref{Fpeta0}) that is relevant, any term in the expansion of the
non-BPS boundary
state (written in terms of the barred variables) can easily be
translated back in terms of the original right moving variables using
the relations (\ref{UbUb}), (\ref{VbVb}) and (\ref{UbVb}). Changing
the definition of the barred variables suitably the non-BPS boundary
state can be given the form of the NS-NS part of any BPS boundary
state discussed in the previous subsection \cite{mukhopadhyay04}.

\sectiono{Computing Physical Quantities}
\label{s:physical}

Given the boundary states in the previous section one would obviously wonder
how to use them to compute physical quantities. It is not
a priori clear how to do such computations. Hoping that the present
method of dealing with boundary states really works, our approach will
be to find prescriptions for such computations.
Below we shall discuss two such issues namely, the closed string
sources and D-brane interaction.

\subsectiono{Sources for Massless Closed String Modes}
\label{ss:one-point}

A D-brane acts like a source for various closed string modes. The
strength of these sources can be computed either by using the boundary
state or the boundary conformal field theory. Without going into much
technical details the NSR computation can be described as follows:
Given a closed string state $|\psi\ra$ there exists a corresponding
conjugate state $\la \psi^{(C)}|$ such that the strength is given by
saturating the conjugate state with the boundary state. The same
result is obtained in the BCFT by computing the disk one-point
function of $\psi^{(C)}$ where the vertex operator is inserted at the
origin of the unit disk. Given a closed string state the correct
conjugate operator has to be found by properly satisfying the ghost
and superghost zero mode saturation rules. Although the details of the
computation should look different in pure spinor formalism, the same
general features are expected to be realized. Here we suggest the
conjugate vertex operators for the massless closed string states and
show how the computation goes through in the present situation. We
begin with the boundary state computation and at the end indicate how
to compute the disk one-point functions. We focus only on the combined
fermionic matter and bosonic ghost part as it is known how to deal
with the rest of the CFT. Let us first define the following operators,
\bea
a^{\mu} = \lam_0\gb^{\mu} \th_0~, &&
\chi_{\al}= (\lam_0\gb^{\mu}\th_0) (\gb_{\mu}\th_0)_{\al}~,
\label{a-chi}
\eea
and similarly for the right moving sector, so that the relevant parts
of the  supergravity states are given by (to the lowest order in
$\theta$-expansion),
\bea
\begin{array}{rrl}
\hbox{NS-NS:} & |V^{\mu \nu}\ra = & a^{\mu} \tilde a^{\nu} |0\ra
\otimes \widetilde{|0\ra}~, \\
\hbox{NS-R:} & |\Psi^{\mu}_{\al}\ra = & a^{\mu} \tilde \chi_{\al}
|0\ra \otimes \widetilde{|0\ra}~, \\
\hbox{R-NS:} & |\bar \Psi_{\al}^{\mu}\ra = & \chi_{\al} \tilde a^{\mu}
|0\ra \otimes \widetilde{|0\ra}~, \\
\hbox{R-R (field strength):} & \quad |F_{\al \bt}\ra = & \chi_{\al}
\tilde \chi_{\bt}
|0\ra \otimes \widetilde{|0\ra}~ . \end{array}
\eea
Then we define,
\bea
a^{(C)}_{\mu} \propto
(\lam_0\gb^{\nu} \th_0) (\lam_0\gb^{\rho}\th_0) (\th_0\gb_{\mu \nu
\rho} \th_0) ~, &&  \chi^{(C)\al} \propto (\lam_0\gb^{\mu}
\th_0) (\lam_0\gb^{\nu}\th_0)(\th_0\gb_{\mu \nu})^{\al}~.
\label{dual-op}
\eea
The proportionality constants are determined by demanding that the
above operators are conjugate to the operators in eqs.(\ref{a-chi}) in the
following sense,
\bea
\la a^{(C)}_{\mu}
a^{\nu} \ra_{Berkovits} = \la \Omega| a^{(C)}_{\mu} a^{\nu} |0\ra =
\dt_{\mu}^{~\nu}~, \quad \la \chi^{(C) \al} \chi_{\bt}
\ra_{Berkovits} = \la \Omega| \chi^{(C)\al}\chi_{\bt} |0\ra =
\dt^{\al}_{~\bt}~.
\label{orthonormal}
\eea
The operators defined in eqs.(\ref{dual-op}) appear at the lowest
orders in $\theta$-expansions of the two elements of BRST cohomology at
ghost number two. The higher order terms are irrelevant for the
present purpose as they do not contribute to the above inner
products. These BRST elements correspond to what are called antifields
associated to gluon ($a^{\mu}$) and gluino ($\chi_{\alpha}$)
respectively\cite{superparticle}. Defining the conjugate states
corresponding to the NS-NS sector in the following way,
\bea
{}_{~I}\la V^{(C)}_{\mu \nu}| =
\widetilde{\la 0|} \tilde a_{\mu} \otimes \la \Omega| a^{(C)}_{\nu} ~,
\label{Vdual}
\eea
we get using form I in eq. (\ref{Fpeta0})\footnote{We follow a
convention where  $\tilde \th^{\al}$ and $\tilde p_{\al}$ pick up a
minus sign while
passing through the spin field corresponding to the state $|0\ra$.},
\bea
{}_{~I}\la V^{(C)}_{\mu \nu}|F,p,\eta\ra_0  =  -\eta_{\mu \rho}
(\mv)^{\rho}_{~\nu}~,
\label{Vdual-Fpeta0}
\eea
which is a desired result. Notice that the right-moving part of the
dual state, namely
$\widetilde{\la 0|} \tilde a_{\mu}$ couples with the term in the
expansion of the boundary state that is linear in both $\tilde w$ and
$\tilde p$ whose left moving part is also linear in both $\lam$ and
$\th$. This particular combination is properly saturated by the left
moving part in the definition (\ref{Vdual}). The overall constant in
the result (\ref{Vdual-Fpeta0}) is not important as it can always be
absorbed in the definition (\ref{Vdual}). Notice that we could also
define the dual state
by interchanging left and right moving objects in eq.(\ref{Vdual}), namely,
\bea
{}_{~II}\la V^{(C)}_{\mu \nu}| = - \widetilde{\la \Omega|} \tilde
a^{(C)}_{\mu} \otimes \la 0|a_{\nu} .
\eea
In that case we need to use form II in eq.(\ref{Fpeta0}) in order to
get the correct result as in eq.(\ref{Vdual-Fpeta0}).

Let us now turn to the R-R states $|F_{\al \bt}\ra$. One would expect
for the corresponding conjugate states to have an index structure:
$\la F^{(C) \al \bt}|$. But since each of the left and right moving
parts of $|F_{\al \bt}\ra$ contributes a spinor index, the conjugate
states, obtained by following the above procedure, have spinor indices
of opposite chirality.
\bea
{}_{~I}\la C^{(C)\al}_{1~~~~\bt}| = \widetilde{\la 0|}\tilde \chi_{\bt}
\otimes \la \Omega | \chi^{(C)\al}~.
\label{C1}
\eea
We would like to interpret these conjugate states to correspond to R-R
potential rather than R-R field strength. Using form I of the boundary
state one can show that the above state produces the following result
which is consistent with this interpretation,
\bea
{}_{~I}\la C^{(C)\al}_{1~~~~\bt}|F,p,\eta\ra_0 = \eta
(\ms)^{\al}_{~\bt} ~. \label{C1-Fpeta0}
\eea
The subscript $1$ in the above notation will now be explained. The R-R
potential has been suggested in the literature to be given by the
following state \cite{cornalba02, grassi04},
\bea
|C_{\al}^{~\bt}\ra = \lt( \chi_{\al}
\lamt^{\bt}_0 + \lam^{\bt}_0 \tilde \chi_{\al} \rt)|0\ra \otimes
\widetilde{|0\ra}~.
\label{C}
\eea
Given this one may wonder why the conjugate state in eq.(\ref{C1}) is
entirely constructed out of $\chi$ and its conjugate. To answer this
question one may consider the following state,
\bea
{}_{~I}\la C^{(C)\al}_{2~~~~\bt}| = \widetilde{\la 0|}\lamt^{\al}_0
\otimes \la \Omega | \lam^{(C)}_{\bt}~,
\label{C2}
\eea
where
\bea
\lam^{(C)}_{\al} \propto (\lam_0\gb^{\mu}\th_0)
(\lam_0\gb^{\nu}\th_0) (\gb^{\rho}\th_0)_{\al}
(\th_0\gb_{\mu\nu\rho}\th_0)~, \hbox{such that } \la
\lam^{(C)}_{\al} \lam^{\bt} \ra_{Berkovits} = \dt_{\al}^{~\bt}~.
\label{lamdual}
\eea
This gives the similar result,
\bea
{}_{~I} \la C^{(C)\al}_{2~~~~\bt} | F,p,\eta \ra_0 = - \eta
(\bar M^{S^T})^{\al}_{~\bt} = \eta ({M^S}^{-1})^{\al}_{~\bt}~.
\eea
We may take the conjugate of the R-R potential in form I to be a linear
combination of ${}_{~I}\la C^{(C)\al}_{1~~~~\bt}|$ and ${}_{~I}\la
C^{(C)\al}_{2~~~~\bt}|$. There are alternative expressions for the
dual states as well which are suitable for computing inner products
with form II of the boundary state. These are given by,
\bea
{}_{~II}\la C^{(C)\al}_{1~~~~\bt}| = \widetilde{\la \Omega|} \tilde
\chi^{(C)\al} \otimes \la 0|\chi_{\bt} ~, && \quad {}_{~II}\la
C^{(C)\al}_{1~~~~\bt}|F,p,\eta
\ra_0 = \eta ({M^S}^{-1})^{\al}_{~\bt}~, \cr && \cr
{}_{~II}\la C^{(C)\al}_{2~~~~\bt}| = \widetilde{\la \Omega|}
\lamt^{(C)}_{\bt} \otimes \la 0|\lam^{\al}_0 ~, && \quad {}_{~II}\la
C^{(C)\al}_{2~~~~\bt}|F,p,\eta \ra_0 = \eta (\ms)^{\al}_{~\bt}~.
\eea

Because of the particular index structure of the R-NS and NS-R states
$|\bar \Psi^{\mu}_{\al}\ra$ and $|\Psi^{\mu}_{\al}\ra$ respectively,
it turns out that for each one of them a conjugate state corresponding
to only one form can be constructed, namely,
\bea
{}_{~I}\la \bar \Psi^{(C) \al}_{~~~\mu}| = \widetilde{\la
0|} \tilde a_{\mu} \otimes \la \Omega | \chi^{(C)\al}~, \quad
{}_{~~II}\la \Psi^{(C) \al}_{~~~\mu}| = \widetilde{\la \Omega |}
\tilde \chi^{(C) \al} \otimes \la 0| a_{\mu}~.
\label{Psidual}
\eea
They can be shown to give zero inner product with the correct form of
the boundary state,
\bea
{}_{~I}\la \bar \Psi^{(C) \al}_{~~~\mu}|F,p,\eta\ra_0 = 0 ~, \quad
{}_{~II}\la \Psi^{(C) \al}_{~~~\mu}|F,p,\eta\ra_0 = 0~,
\label{Psi-Fpeta0}
\eea
which are also expected results.

The dual states used in the boundary state computation should
directly give the vertex operators that need to be used in the
disk one point functions. A list of those operators for the massless
modes is given in table \ref{t:vop}.
\begin{table}[h]
\begin{tabular}{|r|c|c|} \hline
Vertex operator & Form I  & Form II \\
\hline \hline
 && \\
$V^{(C)}_{\mu \nu}(\zeta, \bar \zeta)$  & $a^{(C)}_{\nu}(\zeta)\tilde
a_{\mu}(\bar \zeta)$ & $a_{\nu}(\zeta)
\tilde a^{(C)}_{\mu}(\bar \zeta)$ \\
&& \\
$\Psi^{(C)\al}_{~~~\mu}(\zeta, \bar \zeta)$  & - &
$- a_{\mu}(\zeta) \tilde \chi^{(C)\al}(\bar \zeta)$\\
&&\\
$\bar \Psi^{(C)\al}_{~~~\mu}(\zeta, \bar \zeta)$  &
$\chi^{(C)\al}(\zeta) \tilde a_{\mu}(\bar \zeta)$  & - \\ &&\\
$C^{(C)\al}_{1~~~~\bt}(\zeta, \bar \zeta)$ & $\chi^{(C)\al}(\zeta)
\tilde \chi_{\bt}(\bar \zeta)$ & $\chi_{\bt}(\zeta) \tilde
\chi^{(C)\al}(\bar \zeta) $\\ &&\\
$C^{(C)\al}_{2~~~~\bt}(\zeta, \bar \zeta)$  & $\lam^{(C)}_{\bt}(\zeta)
\tilde \lam^{\al}(\bar \zeta)$ & $\lam^{\al}(\zeta) \tilde
\lam^{(C)}_{\bt}(\bar \zeta)$ \\ \hline
\end{tabular}
\caption{Massless closed string vertex operators to be used in the
computation of disk one-point functions. $(\zeta,
\bar \zeta)$ denote the complex coordinates on a unit disk.}
\label{t:vop}
\end{table}
To compute the disk one-point functions one may first use the
conformal transformation: $z=i(1+\zeta)/(1-\zeta)$ to go from the unit
disk to UHP.
\bea
\la \psi^{(C)}(0,0) \ra^{Disk}_{Berkovits} = \la \psi^{(C)}(i,
-i)\ra^{UHP}_{Berkovits}~,
\label{disk1}
\eea
where $\psi^{(C)}$ is a vertex operator in table \ref{t:vop}. Notice
that all these operators have conformal dimension zero. Given a form I vertex operator one can compute the right hand side of eq.(\ref{disk1}) in the following way:
First define the following operators on the {\it doubled} surface,
\bea
{\cal U}^{\al}(u)=\lt\{\begin{array}{l}
U^{\al}(z)|_{z=u}~, \\ \\
\eta (\ms)^{\al}_{~\bt} \tilde U^{\bt}(\bar z)|_{\zb = u}~,
\end{array}  \rt.
{\cal V}_{\al}(u) = \lt\{ \begin{array}{ll}
V_{\al}(z)|_{z=u}~, & \Im u \geq 0~, \\ &\\
-\eta (\mbs)_{\al}^{~\bt} \tilde V_{\bt}(\bar z)|_{\zb = u}~, & \Im u
\leq 0~,
\end{array}  \rt.
\label{left-doubling}
\eea
then use them to convert the right hand side of eq.(\ref{disk1}) to a
holomorphic correlation function on the full plane\footnote{Notice that although the doubling trick (\ref{left-doubling}) corresponds to boundary conditions (\ref{bcBPS}) in the free CFT, it is being used to compute the correlator (\ref{disk1}) in the constrained theory. This can be justified by the fact that due to covariance (\ref{left-doubling}) gives the correct doubling for any physical operator in the constrained CFT that appears in (\ref{disk1}).}.
For dealing with
the form II vertex operators one follows the same procedure except
that now instead of using the operators in eqs.(\ref{left-doubling})
one uses the following operators,
\bea
\tilde {\cal U}^{\al}(\bar u)=\lt\{\begin{array}{l}
\tilde U^{\al}(\zb)|_{\zb=\bar u}~, \\ \\
- \eta (\bar M^{S^T})^{\al}_{~\bt} U^{\bt}(z)|_{z = \bar u}~,
\end{array}  \rt.
\tilde {\cal V}_{\al}(\bar u) = \lt\{ \begin{array}{ll}
\tilde V_{\al}(\bar z)|_{\zb=\bar u}~, & \Im u \geq 0~, \\ &\\
\eta (M^{s^T})_{\al}^{~\bt} V_{\bt}(z)|_{z = \bar u}~, & \Im u
\leq 0~.
\end{array}  \rt.
\label{right-doubling}
\eea
to convert the right hand side of (\ref{disk1}) to an anti-holomorphic
correlation function. Following this procedure one establishes the
following relation in any relevant form,
\bea
\la \psi^{(C)} |F,p,\eta \ra_0 = \la \psi^{(C)}(0,0) \ra^{Disk}_{Berkovits}~.
\eea

\subsectiono{Comments on Force Computation}\label{s:force}

Interaction between D-branes using boundary states is an important
computation to be
understood in pure spinor formalism. This is also related to the
understanding of
open-closed duality. In NRS formalism we are aware of a simple gauge
choice (Siegel gauge) in which the closed string propagator can be
written in terms
of the world-sheet hamiltonian so that the boundary state when evolved
by this propagator forms the {\it cylinder diagram}. In the present
case the similar gauge choice is not known and therefore it is not
{\it a priory} clear how to go through this computation. But here we
shall try to emphasize yet another feature of the pure spinor
computation. In NSR formalism the associated ghost degrees of freedom
give a certain background independent contribution which cancels two
coordinates (light-cone) worth of contribution from the matter part to
give the correct physical result. From the boundary states discussed
in the previous section it seems difficult to obtain such a background
independent contribution. Below we shall demonstrate this feature by
focusing on the simplest computation of this type namely, the long
range NS-NS force between two parallel BPS D-branes.

Let us first review how the computation goes through in NSR
formalism. The relevant part of the boundary state is the NS-NS part
of a BPS D$p$-brane situated at $\vec y_{\perp}$ along the transverse
directions \cite{NSRbs},
\bea
|NSNS, p, \vec y_{\perp}\ra^{massless}_{NSR} &\propto &\int d\vec
k_{\perp}  e^{-i\vec k_{\perp}.\vec
y_{\perp}} \lt[ \eta_{\mu \rho}
(\mv)^{\rho}_{~\nu} \psi^{\mu}_{-1/2} \tilde \psi^{\nu}_{-1/2} \rt. \cr
&& \lt. - (\g_{-1/2}\tilde \bt_{-1/2} - \bt_{-1/2} \tilde \g_{-1/2} )\rt]
|\vec k_{\perp}, k_{||} =0 \ra_{NSR}~,
\label{NSR-NSNS}
\eea
where the proportionality constant is linear in the tension of the D-brane,
$\psi^{\mu}_{-1/2}$, $\tilde \psi^{\mu}_{-1/2}$, $\bt_{-1/2}$,
$\tilde \bt_{-1/2}$, $\g_{-1/2}$, $\tilde \g_{-1/2}$ are the usual
left and right moving fermionic matter and bosonic superghosts in NS sector
and $|\vec k_{\perp}, k_{||} =0 \ra_{NSR}$ is the ghost number $3$,
picture number $(-1,-1)$  Foch vacuum with the indicated momenta,
\bea
|\vec k_{\perp}, k_{||} =0 \ra_{NSR} = (c_0 + \tilde c_0) c_1 \tilde
c_1 e^{-(\phi + \tilde \phi)}(0,0) e^{i\vec k_{\perp}.\vec X}(0,0)
|0\ra ~.
\label{NSR-vac}
\eea
Notice that the superghost oscillator part in eq.(\ref{NSR-NSNS}) is
the ghost-dilaton which, combined with the trace of
$\psi^{\mu}_{-1/2} \tilde \psi^{\nu}_{-1/2}$, constructs the dilaton state. We
shall see how ghost-dilaton plays its role in producing the correct
physical result for the long distance force. The long distance NS-NS
force between two parallel D-branes D$p^{\prime}$ and D$p$ situated at
$\vec y^{\prime}_{\perp}$ and $\vec y_{\perp}$ is obtained from the
following amplitude,
\bea
{\cal A}_{NSNS} = \int_0^{\infty} dt
{}~~^{massless}_{~~~~NSR}\la NSNS, p^{\prime}, \vec
y^{\prime}_{\perp}| e^{-\pi t (L_0+\tilde L_0)}
|NSNS, p, \vec y_{\perp}\ra^{massless}_{NSR}~,
\label{force1}
\eea
where $L_0$ and $\tilde L_0$ are, as usual, the Virasoro zero
modes. The result turns out to be,
\bea
{\cal A}_{NS-NS} = f(\vec y^{\prime}_{\perp}, \vec y_{\perp}) \lt[ (10-2\nu)
-2\rt]~,
\label{force2}
\eea
where $f(\vec y^{\prime}_{\perp}, \vec y_{\perp})$ is a function whose
details are not needed for our purpose. We are interested only in the
numerical factor kept in the square bracket. $\nu$ is the number of
Neumann-Dirichlet directions involved in this configuration. The
background dependent part $(10-2\nu)$ comes from $\tr(M^{\prime
V}M^{V^T})$ contributed by the fermionic matter part of the state in
eq.(\ref{NSR-NSNS}) whereas the background independent contribution
$-2$ is provided by the ghost-dilaton state. Because of this
contribution the force is zero for a supersymmetric configuration in
which $\nu =4$.

Let us now try to see how the same result might be reproduced in pure
spinor formalism using the boundary state (\ref{bsBPS}). The relevant
part of the boundary state is,
\bea
|NSNS, p, \vec y_{\perp}\ra^{massless}_{PS} &\propto &
{1\over 2} \int d\vec k_{\perp}
e^{-i \vec k_{\perp}.\vec y_{\perp}}\lt[
\exp \lt( - U^{\al T}_0 (\mbs)_{\al}^{~\bt} \sigma_3
\Vt_{\bt,0} \rt) \rt. \cr
&& \lt. + \exp \lt( U^{\al T}_0 (\mbs)_{\al}^{~\bt} \sigma_3
\Vt_{\bt,0} \rt) \rt]
|\vec k_{\perp}, k_{||}=0\ra \otimes |0\ra \otimes
\widetilde{|\hat 0 \ra}~, \cr &&
\label{PS-NSNS}
\eea
Recall that only a finite number of states (which are in one-to-one
correspondence with the physical states) appeared in
eq.(\ref{NSR-NSNS}) for each momentum. But eq.(\ref{PS-NSNS}) contains
an infinite number of states. There are two basic sources for having
lots of extra states in pure spinor formalism. The first one is due to
the fact that the boundary state has been constructed in the gauge
unfixed theory where the $d=9+1$, ${\cal N}=2$ superspace structure
for the target space is manifest.  This results in a lot of auxiliary
fields, which in addition to the physical ones, appear in the boundary
state in (\ref{PS-NSNS}). Secondly, this
boundary state has been constructed in the bigger Hilbert space of the
unconstrained CFT. For the relevant computation one could suggest a
prescription of first projecting the boundary state onto the ones
corresponding to the physical states,
\bea
|NSNS, p, \vec y_{\perp}\ra^{massless}_{PS}
&\rightarrow &  \int d\vec k_{\perp}
e^{-i\vec k_{\perp}.\vec y_{\perp}} \eta_{\mu \rho} (\mv)^{\rho}_{~\nu}
\lt(\lam_0 \gb^{\mu} \th_0 \rt)
\lt(\wt_0 \g^{\nu} \pt_0 \rt) \cr
&&|\vec k_{\perp}, k_{||}=0\ra \otimes
|0\ra \otimes \widetilde{|\hat 0 \ra} ~. \cr &&
\label{PS-NSNS2}
\eea
and then evolving that by world-sheet time evolution. But because of
the absence
of an analog of ghost-dilaton in this formalism one does not get the
background independent contribution of
$-2$ in the square bracket of eq.(\ref{force2}). Absence of the
ghost-dilaton in pure spinor
formalism does not necessarily imply any inconsistency as far as the
present computation is concerned.
This is because we may think of this force computation
simply as a problem in the space-time theory whose spectrum is
correctly reproduced in pure spinor formalism. In this computation one
is simply required to evaluate the tree-level Feynman diagram where
all the NS-NS massless modes propagate between two sources representing
the D-branes. The role of the boundary states is only to provide the
strength of these sources. Extending this observation to  higher
levels one may suggest that the projected boundary state should
actually be thought of as the NSR boundary state in old covariant
quantization which gives correct sources for the space-time fields but
is insufficient to organize the force computation keeping the
open-closed duality manifest.

\sectiono{Discussion}
\label{s:discussion}

We end by mentioning several interesting questions worth exploring in
future work.
\begin{enumerate}
\item
In this paper we constructed D-brane boundary states in the
unconstrained CFT by relaxing the pure spinor constraint. This simply enables
us to work in a larger Hilbert space so that all the inner product
computations relevant to the pure spinor CFT can also be done
here. Therefore these boundary states produce the correct results for
all the closed string sources once the correct prescription is
followed. Despite this advantage these boundary states are not
suitable for computing the cylinder diagram with manifest open-closed
duality. The problem, as explained in the previous section, can not be
solved only by imposing the pure spinor constraint, but the theory
has also to be gauge fixed at all mass levels. In order to compute the 
cylinder diagram with manifest open-closed duality one needs to
project the boundary states constructed here onto a smaller Hilbert
space such that both the pure spinor constraint and the removal of the
gauge degrees of freedom is implemented. This has been explicitly done
in \cite{DDF} by constructing the physical Hilbert space through the
DDF construction. Although this enables us to compute the cylinder
diagram in pure spinor formalism, the projected boundary states are
covariant only under the transverse $SO(8)$ part. Construction of the
boundary states suitable for computing the cylinder diagram with full 
$SO(9,1)$ covariance is still an open question. Computation of the R-R
force between two D-branes is not straightforward even in NSR
formalism. Certain regularization method is involved in this
computation to control the zero modes contribution coming from the
bosonic superghosts \cite{sgh0}. A complete understanding of the
covariant boundary states in pure spinor formalism has to solve all
these subtle issues.
\item
We have also discussed open string boundary conditions in the
unconstrained CFT. Because of the manifest covariance these boundary
conditions produce correct reflection property between the left and
right moving parts of any closed string vertex operator that is
allowed in the pure spinor superstrings. Therefore it should be
possible to use these boundary conditions to compute disk scattering
amplitudes. Here we have prescribed rules for computing disk one point 
functions for massless closed strings only. This needs to be extended to
all possible disk amplitudes with arbitrary number of bulk and
boundary insertions.
\item
The open string boundary conditions for non-BPS D-branes have been
suggested here by generalizing the previous work on light-cone GS
formalism in \cite{mukhopadhyay04}. It should be further investigated if this
generalization gives sensible results. In particular, it would be
interesting to see
if the rules for computing disk amplitudes with arbitrary bulk
insertions, as suggested in \cite{mukhopadhyay04}, also have sensible
generalization to the present case.
\item
In \cite{mukhopadhyay04} it was shown that the bi-local boundary
conditions on space-time fermions for a non-BPS D-brane give rise to
two sectors in the open string spectrum. One is given by periodic
boundary condition resulting in a Bose-Fermi degenerate spectrum same
as that for a BPS D-brane. The other is given by anti-periodic
boundary condition. This sector includes the tachyon and is
responsible for the fermion doubling at the massless level. Following
the same analysis in the pure spinor case also one gets the similar
two sectors. Analysis for the periodic sector goes just like that for
a BPS D-brane. The anti-periodic sector has recently been analyzed in
\cite{DDF} and all the physical open string states have been
explicitly constructed through DDF construction. Despite this
progress it is still unknown how to construct the tachyon vertex
operator. This is because the spin fields of the pure spinor ghosts
with anti-periodic boundary condition are not understood. The same
problem also exists for multiple D-brane configurations at various
angles. In this case the open strings going from one brane to another
will have more general boundary conditions. 
\end{enumerate}

\medskip
\centerline{\bf Acknowledgement}
\noindent
I wish to thank N. Berkovits for helpful communications and
comments on the manuscript. I am also thankful to S. R. Das, G. Mandal,
H. Ooguri, J. Polchinski and A. Shapere for encouraging conversations.
This work was supported by DOE grant DE-FG01-00ER45832.

\appendix

\sectiono{Gamma Matrices and Fiertz Identities} \label{a:gamma}

We denote a $32$-component SO$(9,1)$-spinor in the Weyl basis as
$(\chi^{\al}, \xi_{\al})$, where $\al =1,\cdots , 16$.
The $32$-dimensional gamma matrices are given by,
\bea
\Gamma^{\mu} = \pmatrix{0& \g^{\mu \al \bt} \cr \gb^{\mu}_{\al \bt} & 0}~,
\quad \mu =0,1,\cdots ,9~,
\label{Gamma}
\eea
such that,
\bea
(\g^{\mu}\gb^{\nu} + \g^{\nu}\gb^{\mu})^{\al}_{~\bt} =
2 \eta^{\mu \nu} \dt^{\al}_{~\bt}~, \quad
(\gb^{\mu}\g^{\nu} + \gb^{\nu}\g^{\mu})_{\al}^{~\bt} =
2\eta^{\mu \nu} \dt_{\al}^{~\bt}~.
\label{cliff-g}
\eea
The $16$-dimensional gamma matrices are symmetric and give the following chirality matrix:
\bea
\Gamma = \Gamma^0 \Gamma^1 \cdots \Gamma^9 =
\pmatrix{\g^{01\cdots 9}& 0 \cr 0 & \gb^{01\cdots 9}} =
\pmatrix{\one_{16} & 0 \cr 0 & -\one_{16}}~,
\label{chirality}
\eea
The multi-indexed gamma matrices are defined to be,
\bea
\g^{\mu_1\cdots \mu_{2n} \al}_{~~~~~~~~~\bt} =
(\g^{[\mu_1} \gb^{\mu_2} \cdots \gb^{\mu_{2n}]})^{\al}_{~\bt}~, &\quad &
\gb^{\mu_1\cdots \mu_{2n} ~\bt}_{~~~~~~~~\al} =
(\gb^{[\mu_1} \g^{\mu_2} \cdots \g^{\mu_{2n}]})_{\al}^{~\bt}~, \cr
\g^{\mu_1\cdots \mu_{2n+1} \al \bt} =
(\g^{[\mu_1} \gb^{\mu_2} \cdots \g^{\mu_{2n+1}]})^{\al \bt}~, &\quad &
\gb^{\mu_1\cdots \mu_{2n+1}}_{\al \bt} =
(\gb^{[\mu_1} \g^{\mu_2} \cdots \gb^{\mu_{2n+1}]})_{\al \bt}~,
\eea
The Poincar$\acute{\hbox{e}}$ duality properties are given by,
\bea
\g^{\mu_1\cdots \mu_n} &=&-{1\over (10-n)!} \eps^{\mu_1\cdots \mu_{10}}
\g_{\mu_{n+1}\cdots \mu_{10}} \lt\{ \begin{array}{ll}
(-1)^{n\over 2}~, & n =\hbox{even}~, \\
(-1)^{n+1 \over 2}~, & n=\hbox{odd}~, \end{array} \rt. \cr
\gb^{\mu_1\cdots \mu_n} &=& {1\over (10-n)!} \eps^{\mu_1\cdots \mu_{10}}
\gb_{\mu_{n+1}\cdots \mu_{10}} \lt\{ \begin{array}{ll}
(-1)^{n\over 2}~, & n =\hbox{even}~, \\
(-1)^{n+1 \over 2}~, & n=\hbox{odd}~, \end{array} \rt.
\label{Poincare-dual}
\eea
Various trace formulas are,
\bea
\tr(\g^{\mu_1\cdots \mu_n}) &=& 0 ~, \cr
\tr(\g^{\mu_1\cdots \mu_{2n}} \g^{\nu_1\cdots \nu_m}) &=& 0~,
\quad m\neq 2n~, \cr
\tr(\g^{\mu_1\cdots \mu_{2n+1}} \gb^{\nu_1\cdots \nu_m}) &=& 0~,
\quad m \neq 2n+1~, \cr
\tr(\g^{\mu_1\cdots \mu_{2n}} \g^{\nu_1\cdots \nu_{2n}}) &=&
(-1)^n 16 \D^{[\mu_1\cdots \mu_{2n}], [\nu_1\cdots \nu_{2n}]}~,
\quad n=1,2~, \cr
\tr(\g^{\mu_1\cdots \mu_{2n+1}} \gb^{\nu_1\cdots \nu_{2n+1}}) &=&
(-1)^n 16 \D^{[\mu_1\cdots \mu_{2n+1}], [\nu_1\cdots \nu_{2n+1}]}
+ 16 \dt_{n,2} \eps^{\mu_1\cdots \mu_5 \nu_1 \cdots \nu_5}~,
\quad n=0,1,2~, \cr &&
\label{traces}
\eea
where,
\bea
\D^{[\mu_1\cdots \mu_n], [\nu_1\cdots \nu_n]} &\equiv& \sum_{\cal P}
\hbox{sign}{\cal P} ~\eta^{\{\mu_1\cdots \mu_n \}, {\cal P}\{\nu_1\cdots
\nu_n\}}~, \cr
\eta^{\{\mu_1\cdots \mu_n \}, \{\nu_1\cdots \nu_n\}} &\equiv &
\eta^{\mu_1 \nu_1} \cdots \eta^{\mu_n \nu_n}~.
\label{Delta}
\eea
The sum in the first line goes over $n!$ terms. $\{\mu_1\cdots
\mu_n\}$ is an ordered set and ${\cal P}\{\mu_1 \cdots \mu_n \}$ is
another ordered set obtained by applying the permutation ${\cal P}$ on
$\{\mu_1 \cdots \mu_n \}$. Relations obtained by interchanging $\g$
and $\gb$ matrices in eqs.(\ref{traces}) also hold. Using these
above traces one can prove the following Fiertz identities,
\bea
\chi^{\al} \xi^{\bt} &=&
{1\over 16} (\chi\gb^{\mu}\xi)\g_{\mu}^{\al \bt} +
{1\over 16\times 3!} (\chi \gb^{\mu_1\cdots \mu_3}\xi)
\g_{\mu_1\cdots \mu_3}^{\al \bt} +
{1\over 16 \times 5! \times 2} (\chi \gb^{\mu_1\cdots \mu_5}\xi)
\g_{\mu_1\cdots \mu_5}^{\al \bt}~, \cr
\chi_{\al} \xi_{\bt} &=&
{1\over 16} (\chi\g_{\mu}\xi)\gb^{\mu}_{\al \bt} +
{1\over 16\times 3!} (\chi \g_{\mu_1\cdots \mu_3}\xi)
\gb^{\mu_1\cdots \mu_3}_{\al \bt} +
{1\over 16 \times 5! \times 2} (\chi \g_{\mu_1\cdots \mu_5}\xi)
\gb^{\mu_1\cdots \mu_5}_{\al \bt}~, \cr
\chi^{\al}\xi_{\bt} &=& {1\over 16} (\chi \xi) \dt^{\al}_{~\bt} +
{1\over 16\times 2!}(\chi\gb^{\mu_1\mu_2}\xi) \g_{\mu_1\mu_2~\bt}^{~~~~\al}
+{1\over 16\times 4!}(\chi\gb^{\mu_1\cdots \mu_4}\xi)
\g_{\mu_1\cdots \mu_4~\bt}^{~~~~~~\al} ~, \cr
\chi_{\al}\xi^{\bt} &=& {1\over 16} (\chi \xi) \dt_{\al}^{~\bt} +
{1\over 16\times 2!}(\chi\g^{\mu_1\mu_2}\xi) \gb_{\mu_1\mu_2 \al}^{~~~~~~\bt}
+{1\over 16\times 4!}(\chi\g^{\mu_1\cdots \mu_4}\xi)
\gb_{\mu_1\cdots \mu_4 \al}^{~~~~~~~~\bt} ~.
\label{fiertz}
\eea

\end{document}